\documentclass[%
 amsmath,amssymb,
 aps, physrev,
]{revtex4-2}

\usepackage{graphicx}
\usepackage{dcolumn}
\usepackage{bm}

\usepackage{multirow}
\usepackage[acronym]{glossaries}
\makeglossaries
\usepackage{tikz}
\usepackage[european]{circuitikz}
\usetikzlibrary{external}
\newacronym{IVU}{IVU}{In-vacuum undulator}
\newacronym{RF}{RF}{radio frequency}
\newacronym{RFPG}{RFPG}{RF particle gun}
\newacronym{3D EM}{3D EM}{3D electromagnetic}
\newacronym{EM}{EM}{electromagnetic}
\newacronym{EQC}{EQC}{equivalent circuit}
\newacronym{RWG}{RWG}{rectangular waveguide}
\newacronym{WG}{WG}{waveguide}
\newacronym{VSWR}{VSWR}{Voltage standing wave ratio}

\begin{document}

\title{Equivalent circuit modelling as a design tool for particle accelerator RF components}

\author{Dieter Pelz}
\email{Contact author: dpelz@rfcurrent.com}
 \affiliation{RFCurrent, Melbourne, Australia}
 
\author{Mark J. Boland}%
\affiliation{%
 Department of Physics and Engineering Physics, University of Saskatchewan, Saskatoon, Canada.
}%


\begin{abstract}
This paper presents a comprehensive approach to generating and utilizing accurate equivalent circuits for structures commonly found in the particle accelerator domain. After a discussion of equivalent circuits in general, a systematic process for determining the appropriate circuit topology is outlined, and a method for finding circuit element values is introduced. The proposed methodology is demonstrated in detail, supported by four illustrative examples. Equivalent circuit element values are provided. An example of a decomposition strategy for handling complex structures is given. Additionally, a list of typical applications of equivalent circuits is provided with brief descriptions.
\end{abstract}

\maketitle

\section{\label{sec:intro}Introduction}

In our recent paper on the analysis and removal of unwanted \gls{RF} resonances in an \gls{IVU}~\cite{Ericson2025}, an equivalent circuit for a \gls{IVU} was presented. 
It was generated for the purpose of confirmation of the mode-nature of critical resonances. 
This article was written in response to a suggestion to devote a separate paper to the derivation and generation of equivalent circuits of  structures typically used in the field of particle acceleration.

Equivalent circuit models are useful tools in connection with \gls{3D EM} simulation. 
They are widely used in the field of \gls{RF} Engineering~\cite{Matthaei1964,Marcuvitz1951,Levy2000, Cameron2007} for design purposes and for diagnostics. 
Well established design methods, like \emph{Co-simulation} and \emph{Space Mapping} techniques both rely heavily on circuit equivalence~\cite{Bandler1994}.

While \gls{3D EM} simulation of $n$-port structures provides us with valuable \gls{EM} field data, surface currents, port $S$-parameters, etc., it does not directly tell us much about the internal nature of the simulated structure. 
This is where a circuit model becomes meaningful and important. 
Clearly, the better we understand the inner workings of a structure, the better we can design it and optimise its performance~\cite{Levy2000,Apsimon2017}. The power of this clarity of a design is what we aim to highlight in this paper.

A certain amount of detailed information on generating equivalent circuits is being provided in this article. 
It is however not intended to be a complete instruction. 
Instead, the aim is to show that it is relatively easy to generate equivalent circuits for use as a powerful diagnostic tool and as a design process enhancing tool.

\section{Equivalent circuits}
Historically, equivalent high-frequency circuits became a necessity when higher-frequencies ($> 300$~MHz) were used in the first half of the last century. 
These circuits were often used in conjunction with uniform transmission line elements for analysis and design work to overcome the problem that existed at junctions and other discontinuities. 
A classic example would be a step in the inner conductor of a coaxial line~\cite{Marcuvitz1951}. 
Lumped element equivalent circuits were commonly used for the entire design process \cite{Matthaei1964} until 3D EM simulation software became widely available around the year 1997.

Equivalent circuits are found in many particle accelerator physics text books. 
A typical example is the representation of an accelerator cavity as a parallel-tuned $LC$ circuit~\cite{Wiedemann2015}. 
With few exceptions, these text book circuit models are mostly of qualitative nature insofar as there are neither $LC$ element values given, nor a calculation method provided. 

Several good articles exist in which equivalent circuit techniques in connection with particle acceleration are used \cite{Vretenar2013,Turner1992,Apsimon2017,Li2024,Hu2022,Vaughn2024,Whiting1968}. 
It can also be said that the so-called "fast frequency sweep" in modern 3D EM simulators (such as CST \cite{CST}, HFSS \cite{HFSS}) is based on an equivalence principle.
\\
Quantitative equivalence of an \gls{EQC} may be related to port $S$-parameter equivalence in the frequency domain or to other properties for which a valid circuit model exists. 
An \gls{EQC} model that yields the same $n$-port $S$-parameters as those obtained from \gls{3D EM} simulation over a given frequency range, is a valid \gls{EQC}. 
Mathematically, this leads us to requiring that:
\begin{equation}
\label{eqn: S_EM = S_EQC}
\left( S \right)_{\text{EM}} = \left( S \right)_{\text{EQC}}.
\end{equation} 
While totally different in nature, both the 3D EM simulation and the simulation of a valid equivalent circuit yield the same port $S$-parameters over a given frequency range. 
Certain equivalent circuit elements may even have a strong relation to localised  parts  of the given 3D structure, like for example coupling is usually related to mutual distance or to the size of a coupling-related 3D sub-structure. 
A relevant example is the diameter of a circular iris in a linear accelerator which determines the \gls{RF} coupling between cavities. 

The achievable accuracy or the quality of circuit equivalence is usually so high, that it is nearly impossible to distinguish between response graphs from 3D EM simulation and equivalent circuit simulation.

Circuit equivalence is by its own nature frequency band limited.
Strictly, we should not use an equivalent circuit outside its range of validity. 
The latter is usually given by the used \gls{3D EM} simulation result data. 
Validity may not always deteriorate rapidly when going beyond its limits, but care must be taken when doing so. 
Some examples of equivalent circuits will be presented in this paper.

\section{Generating an equivalent\\circuit}
\label{Generating EQC}
For relatively simple structures, like TEM-mode resonators, there are analytical equations which directly yield the equivalent circuit element values~\cite{Collins2001,Matthaei1964,Sarbacher1943}. 
For uniform transmission lines (constant cross-sectional dimensions) analytical circuit elements exist as well. 
Complex structures, however, require a more elaborate approach, which can be subdivided into the following steps:
\begin{enumerate}
    \item Visual inspection of \gls{3D EM} model structure and observation of EM field distribution (field plots and field animation); 
    \item Decomposition of \gls{3D EM} model into substructures if required;
    \item Composition of candidate equivalent circuit;
    \item \label{item:init} Setting of initial circuit element values (analytical, prior knowledge, experience);
    \item Setting of constraints for element value ranges (usually a small percentage of initial value);
    \item Export of \gls{3D EM} $S$-parameter data and import into circuit simulator;
    \item \label{item:circsim} Circuit simulation and comparison with \gls{3D EM} simulation $S$-parameter data;
    \item \label{item:iterate} Iterative manual refinement of element values (observation, systematic trials, adjustment);
    \item Definition of goals for minimisation of an error-function value Eq.~(\ref{eqn: S_EM = S_EQC}), Eq.~(\ref{eqn: goal S_ij});
    \item Minimisation of error function value (observation of convergence, adjustment of element value constraints, refinement);
    \item \label{item:update} Updating of circuit model element values with improved values;
    \item Repetition of items \ref{item:init} to \ref{item:update} until a minimum has been reached and the error function value is as low as desired.
\end{enumerate}

The process of minimisation of an error function between two sets of data is also referred to as \emph{fitting} by which is meant the adjustment of circuit element values with reference to the simulation data set of another model, usually \gls{3D EM}.
With reference to items \ref{item:circsim} and \ref{item:iterate} in the above list it is usually recommendable for $n$-port structures to focus first on the transmission $S$-parameters in terms of comparing \gls{3D EM} and \gls{EQC} $S$-parameter responses while making adjustments in the circuit model (and also when starting the minimisation process). 
Once a level of agreement between the two has been established, the reflection $S$-parameters can be included for further refinement of the \gls{EQC}.

The number of frequency points used for the circuit simulation during the minimisation process should be adjusted to a minimum for reasons of good convergence of the minimisation. The number of points should, however, be large enough to ensure that all features of the response graphs are captured, i.e. sampled. 

When a good synchronicity between the \gls{3D EM} model and the equivalent circuit has been achieved, the resulting \gls{EQC} element values offer clues about the structure. In certain cases one may want to perform
a fast optimisation of the equivalent circuit in terms of desired performance criteria in order to obtain at the least a direction for optimising the \gls{3D EM} model. 
The information obtained from the process is generally of very high diagnostic and educational value. One can for example make a small change in the 3D model structure and then re-synchronise with and use the \gls{EQC} to find out more about the effects of the change made (see Section~\ref{App}).

The work effort involved in creating an equivalent circuit may seem considerable but so are the lasting benefits of having a valid equivalent circuit \cite{Levy2000}, \cite{Apsimon2017}. 

\subsection{Finding the EQC topology}
\label{sec:finding}
The EQC topology can usually be derived from the 3D model by visual inspection in conjunction with observation of \gls{EM} field patterns. Here, the field vector plots typically serve best. We refer to the examples in Section \ref{sec:examples} that explain and illustrate this process in more detail. Most structures can be sub-divided into basic elements like uniform transmission lines, cavities, couplings, etc. In many cases, an iterative process may be required, where, for example, the S-parameter response graphs of a candidate \gls{EQC} are compared with those of \gls{3D EM} simulation. Once a certain qualitative agreement (preservation of features of a frequency response) has been achieved, a refinement process for the \gls{EQC} element values may be started. 

\subsection{Refinement of equivalent circuit element values by mathematical minimisation}
\label{sec:refine}

Most of the work involved in generating an equivalent circuit can be done on a circuit simulator. 
Modern circuit simulators~\cite{ADS,AWR} provide a range of so-called optimisers which usually are applied for the optimisation of a desired circuit performance criterion. 
In our case the goal is essentially to minimise the difference between the simulation data sets obtained from \gls{3D EM} simulation and those obtained from simulation of a candidate \gls{EQC} (Eq.~(\ref{eqn: S_EM = S_EQC})).
Given that we start with such circuit element values that are near to the final values, the gradient optimiser is usually the best choice.
A typical goal in this case could be the minimisation of the difference of the transmission coefficients $S_{ij}$. 
We need to define the goal on the basis of Eq.~(\ref{eqn: S_EM = S_EQC}). 
For numerical reasons it is however advisable to define a relative difference of the complex $S$-parameters as:
\begin{equation}
\label{eqn: goal S_ij}
\frac{|S_{ij\_\text{EM}} - S_{ij\_\text{EQC}}|}{|S_{ij\_\text{EM}}|}
\end{equation}
By using Eq.~(\ref{eqn: goal S_ij}) in the minimization goal function we ensure that the circuit equivalence quality extends uniformly over the entire $S$-parameter value range. 
Equation~(\ref{eqn: goal S_ij}) is evaluated over a given number of frequency points and least-squares summed up to form a so-called error function value. (The terms in the numerator difference (\ref{eqn: goal S_ij}) are of course the complex $S$-parameters whereas the denominator is an $S$-parameter magnitude.)
Mathematically, we can formulate the process of finding the values of the equivalent circuit elements as follows.
%
\begin{equation}
\label{eqn: x_i eqn general}
x = \arg\min\frac{|(S)_{_\text{EM}} - (S)_{_\text{EQC}}(x)|}{|(S)_{_\text{EM}}|}
\end{equation}
where $x$ is a vector of the equivalent circuit parameter values which can be identified as the \emph{arguments} of a hypothetical minimisation function \cite{Bandler1994}.
In practice, Eq.~(\ref{eqn: x_i eqn general}) is accomplished by using standard functions of a suitable circuit simulator~\cite{ADS,AWR}. 

The minimisation process requires observation of convergence and occasionally also a certain amount of mindful re-setting of circuit element value constraints before re-starting the process. 
The process ends when a minimum with a low residual error value is found and all variable values are well within the range given by their numerical constraints. The process of finding the equivalent circuit values is also often referred to as \emph{extraction}. Monitoring of response graphs during the minimisation is advisable. In certain situations, manual interruption may be necessary. In such cases, further manual refinement of circuit element values may be required. 

It should be noted that passive circuit analysis is a relatively simple matter and various analysis methods exist~\cite{Chua1975}. 
Among them, the so-called \emph{nodal admittance matrix} analysis is perhaps the easiest and also the most convenient one for passive $RLC$-circuits. 
The definition of the required matrix elements follows directly from the elements and the topology of the node-numbered circuit model. 
Gradient type optimisation subroutines can also be integrated. 
Therefore, implementation in popular maths environments, like Matlab~\cite{Matlab} or MathCAD~\cite{Mathcad}, is relatively easy. 

\section{Examples of equivalent\\circuits}
\label{sec:examples}
The following examples are intended for further explanation and illustration of the process of generating equivalent circuits. 
The chosen example structures are not to be understood as final designs for any given purpose. 
They merely serve as example structures for the purpose of demonstrating how equivalent circuits are derived and generated.

\subsection{RF-cavity particle gun equivalent circuit}
\label{RFPG}

\subsubsection{General description}

\begin{figure}[hb]
    \centering
    \includegraphics[width=.8\linewidth,angle=-90,trim=110 10 100 0,clip]{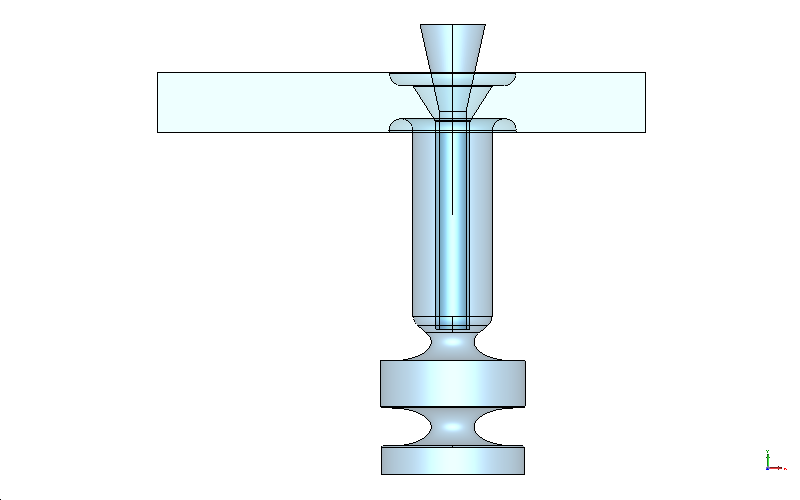}
    \caption{Cut view of RFPG where the particle beam travels from left-to-right along a central axis.}
    \label{fig:cut_view_of_RFPG}
\end{figure}

As the first example, the derivation of an accurate \gls{EQC} for an \gls{RFPG}, consisting of an \gls{RF}-cavity and an input coupler, shall be presented. 
The narrowband nature of the \gls{RFPG} lends this structure to developing an \gls{EQC}. The overall \gls{RFPG} \gls{3D EM} model structure is shown in Fig.~\ref{fig:cut_view_of_RFPG}. 

It consists of a waveguide-to-coaxial transition, often casually referred to as \emph{door-knob waveguide coupler}, and a coaxially fed particle gun. 
A \gls{WG} section with a port is used for feeding \gls{RF} power followed by a \gls{WG}-to-coaxial transition.  An adjustable end-shorted \gls{WG} stub is connected next to the \gls{WG}-to-coaxial transition for the purpose of adjustment of the input VSWR.  

The coaxial line connected to the \gls{WG}-to-coaxial transition guides the \gls{RF} power to the cavities via E-field coupling. Its inner tube is also a beam-tube and therefore doubly-utilised. E-field iris coupling exists between the cavities. The cathode side cavity is a \emph{half-length cavity} whereas the coax-side cavity is \emph{full-length}.

\subsubsection{Subdividing the structure}
The overall structure is considerably complex in the context of an \gls{EQC}. Therefore, subdivision of the \gls{RFPG} structure shall be considered for making it easier to generate a valid and accurate \gls{EQC}. 
\begin{figure}[ht!]
    \centering
    \includegraphics[scale=0.3,angle=-90]{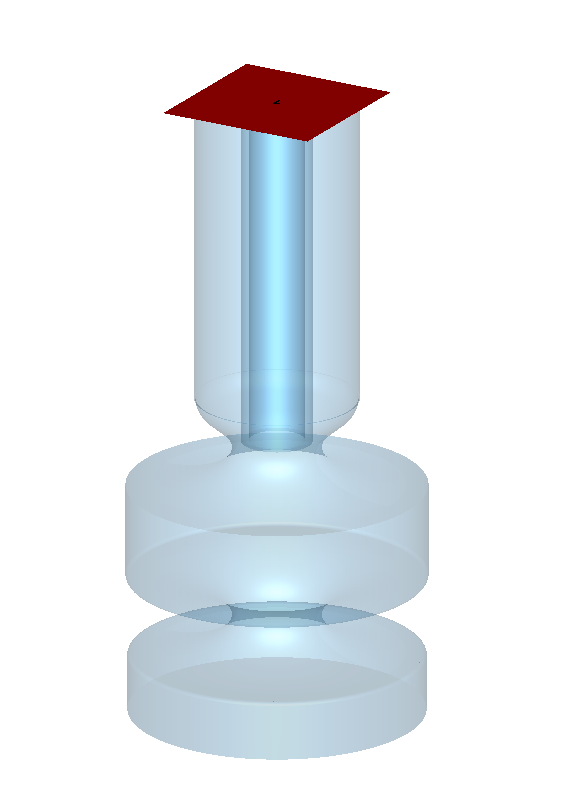}
    \caption{RFPG sub-structure with coaxial port cut in red and showing the rotational symmerty about the beam axis.}
    \label{fig: RFPG sub}
\end{figure}
\begin{figure*}[ht!]
\centering
\includegraphics[width=0.5\textwidth]{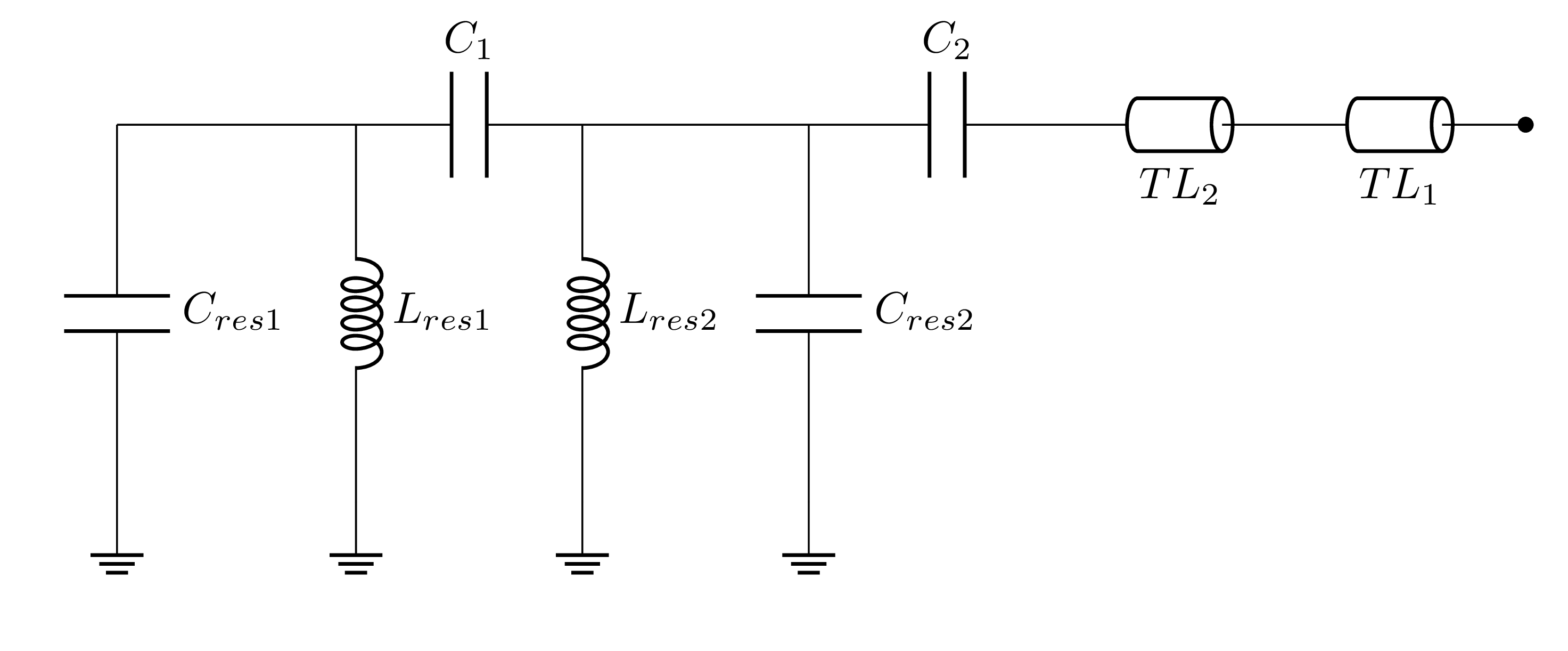}
\caption{EQC of sub-structure shown in Fig.~\ref{fig: RFPG sub}.}
\label{fig: EQC of sub}
\end{figure*}

The coaxial line from the \gls{WG}-to-coaxial transition to the cavities is relatively long and for the purpose of simplification, the overall structure can be split into two parts by \emph{cutting} through the coaxial line and placing an $S$-parameter ports on the cut cross-section (see Fig.~\ref{fig: RFPG sub}). 
It is fully legitimate and proven to place $S$-parameter ports on the now-exposed end of the coaxial line sections (see Fig.~\ref{fig: EQC of sub}). 
The distance of the port-plane to the \gls{WG} must be chosen so that a pure TEM-mode exists at the port-plane. 
This requires keeping a sufficient distance from the \gls{WG} transition (door-knob) where the fields have not yet settled to a pure TEM-mode. 

\subsubsection{\gls{RFPG} equivalent circuit composition}
With reference to the right side of Fig.~\ref{fig: RFPG sub} it is clear, that the first equivalent circuit element must be a uniform coaxial transmission line ($TL_1$ in Fig.~\ref{fig: EQC of sub}) of known characteristic impedance. As for its length we only know the maximum length which is the length of the inner conductor, which doubles as a beam tube. At the cavity end of the coaxial \emph{end-open} tube, the diameter of the outer conductor tapers with a certain curvature. It is therefore reasonable to represent at least a section at the open-end of the coaxial line by a short piece of coaxial line ($TL_2$) with a lower characteristic impedance to be refined later. The sum of the lengths of the two transmission lines $TL_1$, $TL_2$ must be close to the physical length of the inner tube conductor (TEM-mode), assuming that the loading by the cavity via $C_2$ is very small. In this context it is important to see that the mechanical line lengths of the 3D model are not necessarily identical to the so-called \emph{effective electrical lengths} used in the \gls{EQC}.

Knowing that the E-field (e.g. from inspection of \gls{3D EM} model field results) at the cavity end of the inner tube must be coupling to the first cavity, we are able to represent such coupling by a series capacitor $C_2$ of relatively small value to be subsequently refined. The cavities can be fully represented by parallel-tuned $LC$ circuits with the $Q$-factor embodied by a resistor in series with the resonator inductance (not shown). As an educated guess, we assume that the $LC$ circuits are shunt connected to ground. The reason is that the coupling to and from the cavities is located at the $E$-field maximum of the $TM_{010}$-mode cavities.

It is sufficient for the initial circuit element values to be quite rough estimates. Often, the typical value ranges are known from experience and from literature. For resonant circuits in the equivalent circuit, the resonant frequency is often known (here: from \gls{3D EM} $S_{11}$ results) and this enables forcing the correct resonance in the equivalent circuit by setting a value for $L$ or $C$ and then calculating the other element value from the resonance condition at the given resonant frequency. For the case of a set  capacitance, the inductance is calculated as:
\begin{equation}
\label{eqn: L calc}
L=\frac{1}{\omega_0^2 C_{\text{node}}}
\end{equation}
It has to be noted here that the resonance in question is that of the given network node to which the $LC$ resonator is connected. By this we mean that all other connected branches which may affect the nodal resonance must be considered (see nodal admittance matrix theory \cite{Chua1975}). In the given case (Fig.~\ref{fig: EQC of sub}) this simply means that $C_1$ needs to be added to $C_{\text{res1}}$ and $C_1$ and $C_2$ are to be added to $C_{\text{res2}}$.
While this may seem trivial, it is an important detail which ensures that the minimisation process converges fast and does not unnecessarily alter the known resonant frequencies excessively. 

The starting values of the unloaded $Q$-factors can be taken as those of a closed cavity for which analytical equations exist, reduced by a small percentage to account for additional cavity losses. The material conductivity of the actual cavity (\gls{3D EM} model) must be used. 

Further iterative manual refinement of the initial circuit element values typically converges fast and it is thereby relatively easy to get near enough to the actual values. 

\begin{figure}[ht]
    \centering
    \includegraphics[width=1\linewidth]{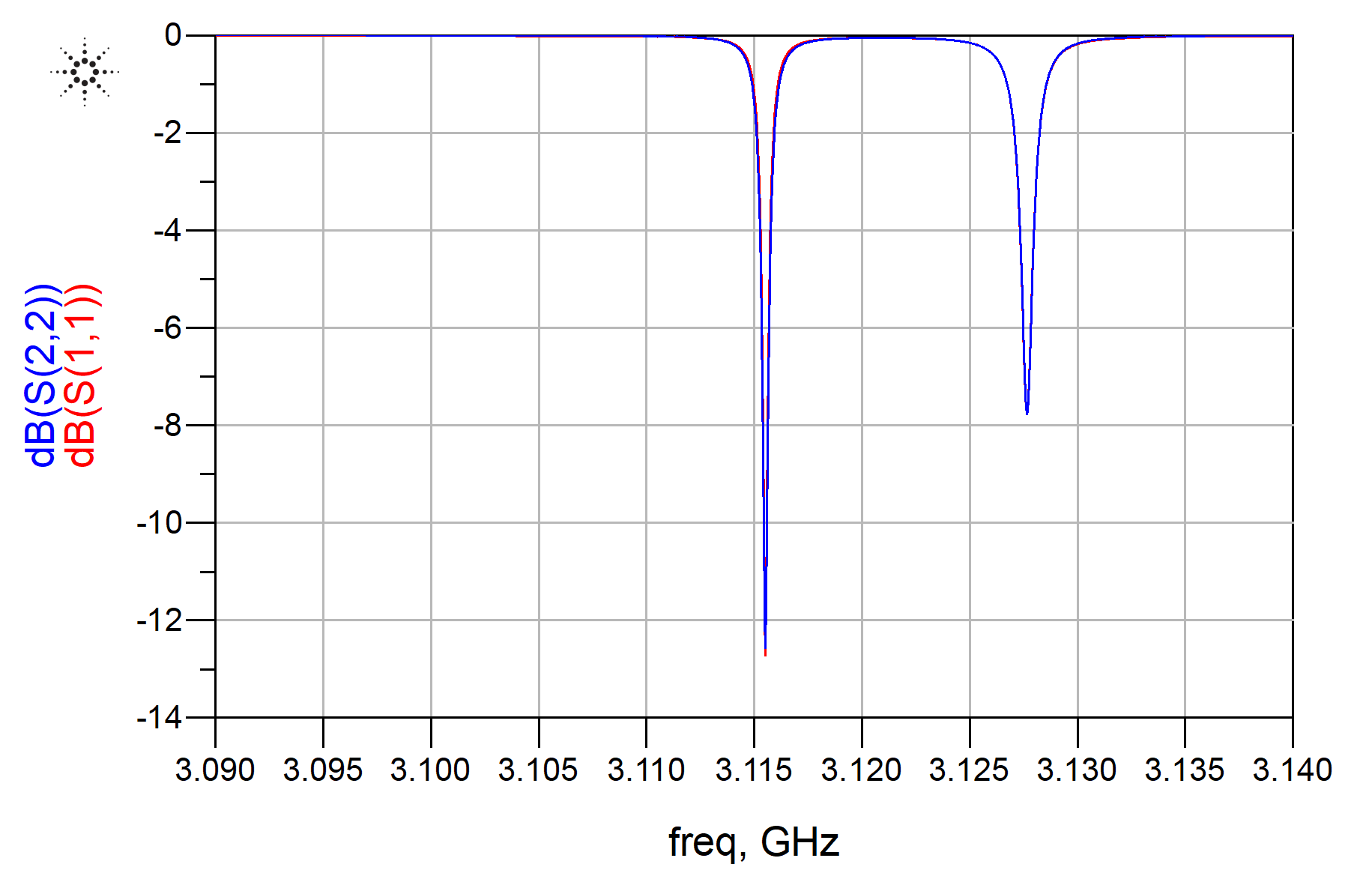}
    \caption{Equivalent circuit of RFPG sub-structure and \gls{3D EM} simulation results. $S_{11}$ (red): \gls{3D EM}, $S_{22}$ (blue): \gls{EQC}. The blue curve is very nearly congruent with the red curve.}
    \label{fig: RFPG graph EQC and 3D EM}
\end{figure}

With careful preparation, the succeeding minimisation process using Eq.~(\ref{eqn: goal S_ij}) usually converges rapidly. The 1-port nature of the structure limits us to use a $S_{11}$ \gls{3D EM} data set. Depending on how the variable constraints are set, the minimisation may need to be re-started with adjusted element value constraints enabling the process to continue. Typical final least-squares error values are in the order of 0.001 to 0.00001. 
Fig.~\ref{fig: RFPG graph EQC and 3D EM} shows the rather good agreement between results obtained from simulations of the \gls{EQC} and the \gls{3D EM} model.

With reference to Fig.~\ref{fig: EQC of sub} the final circuit element values are listed in Table~\ref{tab:res_param}.
%
\begin{table}[ht!]
\caption{List of circuit elements values for the EQC in Fig.~\ref{fig: EQC of sub}.}
\label{tab:res_param}
\centering
\begin{tabular}{|l|c|c|}
\hline
\textbf{Parameter} & \textbf{Value} & \textbf{Unit} \\
\hline
$f_{\text{res1}}$ & 3118.33 & MHz \\
$Q_{\text{res1}}$ & 6481 & -- \\
$C_{\text{res1}}$ & 4.6226 & pF \\
$L_{\text{res1}}$ & 0.5633 & nH \\
$C_1$ & 0.02225 & pF \\
\hline
$f_{\text{res2}}$ & 3124.58 & MHz \\
$Q_{\text{res2}}$ & 11879 & -- \\
$C_{\text{res2}}$ & 9.731 & pF \\
$L_{\text{res2}}$ & 0.2644 & nH \\
$C_2$ & 0.078 & pF \\
\hline
$TL_1: Z_0$ & 48.5 & Ohm \\
$TL_1: l_1$ & 88.09 & mm \\
$TL_2: Z_0$ & 31.44 & Ohm \\
$TL_2: l_2$ & 14.26 & mm \\
\hline
\end{tabular}
\end{table}

The inductance values of the resonant circuits are obtained via the given resonant frequencies and the nodal capacitances as described earlier in Eq.~(\ref{eqn: L calc}). For the identification of the resonators it may be necessary to slightly detune one of the cavities in the \gls{3D EM} model. This can be done by way of so-called \emph{perturbation} in the form of a small metal object (e.g. a disc) inside the chosen cavity. A subsequent simulation will then show which of the two resonances has moved and thus will identify the resonator.

\subsubsection{Waveguide to coax transition equivalent circuit}
\begin{figure}[ht]
    \centering
    \includegraphics[scale=0.2,angle=-90]{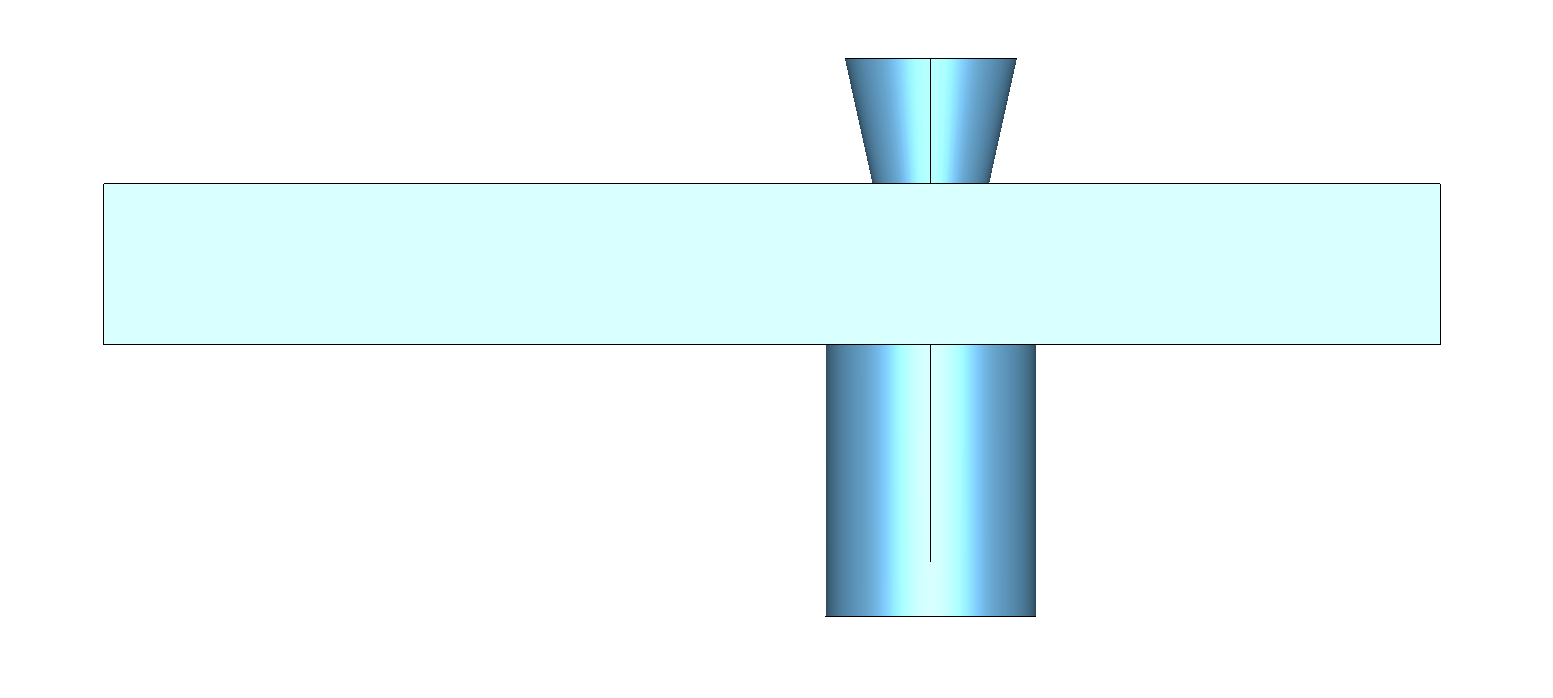}
    \caption{RFPG \gls{WG} to coax sub-structure.}
    \label{fig: RFPG WG coax}
\end{figure}

\begin{figure*}[hb!]
\centering
\includegraphics[width=0.5\textwidth]{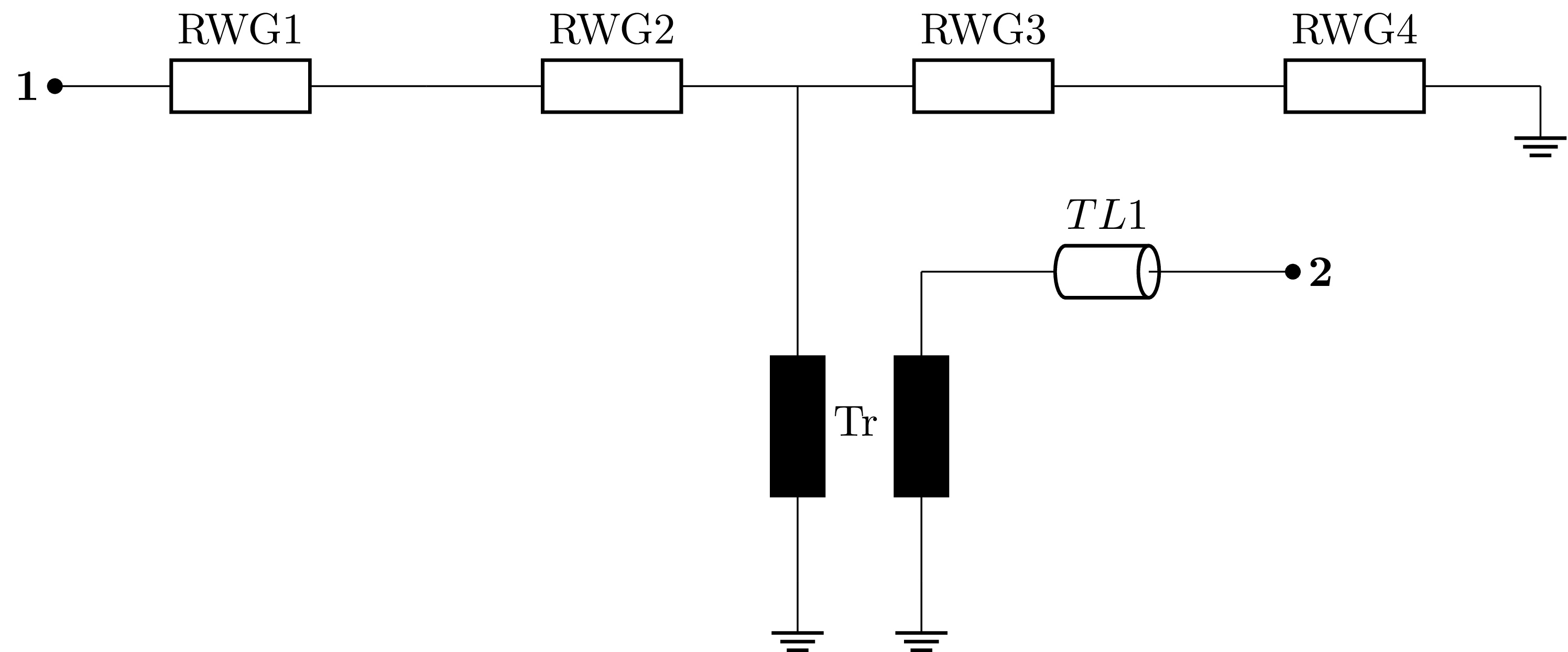}
    
\caption{RFPG EQC of \gls{WG} to Coax substructure.}
\label{fig: RFPG eqc WG to coax}
\end{figure*}

For the sake of completeness, a valid equivalent circuit (see Fig.~\ref{fig: RFPG eqc WG to coax}) was also created in the same manner for the \gls{WG} to coax transition shown in Fig.~\ref{fig: RFPG WG coax}. \gls{WG} circuit elements available in the circuit simulator were used. Given the relatively narrow simulation frequency range, WG dispersion could also be neglected and simple transmission line elements (as used in Fig.~\ref{fig: EQC of sub}) would probably also be sufficient. 

\begin{figure}[ht]
    \centering
    \includegraphics[width=1\linewidth]{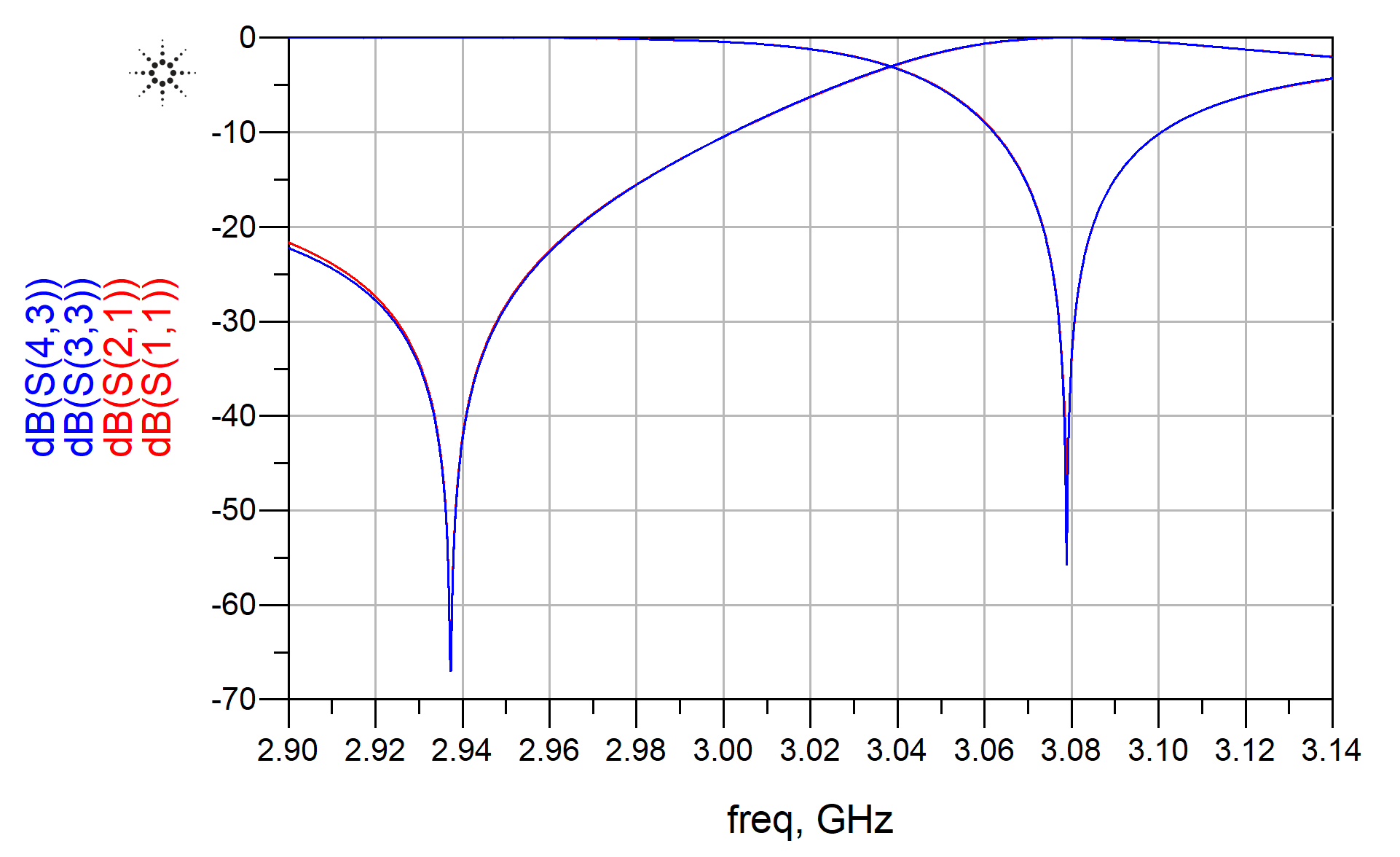}
    \caption{\gls{WG} to coax transition equivalent circuit and 3D EM simulation results. The blue curves are very nearly congruent with the red curves.}
    \label{fig: RFPG WG_coax Spar graph}
\end{figure}

The discontinuities presented by the \gls{WG} to coax transition required to be represented by sections of lower WG impedance (RWG 2 and 3).
The equivalent circuit contains an ideal transformer to represent the involved impedance transformation to suit the connected coaxial line. This was found sufficient in terms of good equivalence of the circuit model. Good agreement between \gls{3D EM} and equivalent circuit simulation results is illustrated in Fig.~\ref{fig: RFPG WG_coax Spar graph}. This is perhaps a good example for a relatively easy way of creating  circuit equivalence. 

The equivalent circuits shown in Fig.~\ref{fig: EQC of sub} and Fig.~\ref{fig: RFPG eqc WG to coax} can be combined to form an equivalent circuit for the entire structure shown in Fig.~\ref{fig:cut_view_of_RFPG}. 

With reference to Fig.~\ref{fig: RFPG eqc WG to coax}, the equivalent circuit element values are shown in Table~\ref{tab:wg_params}.
\begin{table}[ht!]
\caption{Rectangular Waveguide (RWG), Transmission Line (TL), and Port Parameters for the RFPG EQC in Fig.~\ref{fig: RFPG eqc WG to coax}.}
\label{tab:wg_params}
\centering
\begin{tabular}{|l|c|c|c|}
\hline
\textbf{Comp.} & \textbf{A (mm)} & \textbf{B (mm)} & \textbf{Length (mm)} \\
\hline
RWG1 & 72.14 & 33.64 & 85.17 \\
RWG2 & 72.14 & 18.62 & 40.79 \\
RWG3 & 72.14 & 18.62 & 49.59 \\
RWG4 & 72.14 & 33.64 & 142.51 \\
\hline
\end{tabular}

\vspace{0.5cm}

\begin{tabular}{|l|c|c|}
\hline
\textbf{Component} & \textbf{Parameter} & \textbf{Value} \\
\hline
Tr & $n$ & 1.86 \\
TL1 & $Z$ (Ohm) & 48.47 \\
TL1 & Length (mm) & 64.45 \\
P1 & $Z$ (Ohm) & 493.5 \\
P2 & $Z$ (Ohm) & 48.47 \\
\hline
\end{tabular}
\end{table}

\subsubsection{Additional proof of correctness of \gls{RFPG} \gls{EQC} in Fig.~\ref{fig: EQC of sub}}
In addition to the above, we also inserted a very weak coupling probe into the outer wall of the cathode-side cavity in the \gls{3D EM} model. This probe created a 2nd port enabling us to also obtain a $S_{21}$ transmission $S$-parameter data. The probe was represented by a small coupling capacitance at the cathode side resonator node ($C_{\text{res1}}$, $L_{\text{res1}}$ in Fig.~\ref{fig: EQC of sub}) in the equivalent circuit. As expected, the minimisation process led to a near-perfect agreement between \gls{3D EM} and equivalent circuit $S_{21}$ response. Only the resonator capacitance showed a minor change, caused by the weak coupling probe. 

\subsection{In-vacuum undulator (IVU) equivalent circuit}
To avoid unnecessary duplication, we ask readers to refer to Ref.~\cite{Ericson2025} for the overall IVU structure and other details.

\subsubsection{Identity of IVU resonances}

Harmful IVU resonances in the VHF frequency range were identified via \gls{3D EM} Eigenmode simulation as TEM-mode resonances. These resonances exhibit a strong relationship to the length of the IVU~\cite{Ericson2025}. Independent confirmation of the nature of the resonances required that an equivalent circuit was created. 
The equivalence criteria of our equivalent circuit were the frequencies of the IVU resonances.
For finding the element values of the \gls{EQC} we therefore demanded that
\begin{equation}
\label{eqn fres}
f_{{\text{resEM}}_i} = f_{{\text{resEQC}}_i}\hspace{1cm}  i = {\text{1 ... n}}
\end{equation}
Uniform transmission line elements used in the equivalent circuit (Fig.~\ref{fig: IVU_EQC}) are only valid for the TEM-mode. An \gls{EQC} that exhibits the same resonances as does the \gls{3D EM} simulation model would therefore confirm beyond reasonable doubt that the critical IVU Eigenmodes obtained from \gls{3D EM} simulation are indeed of TEM-mode nature. The TEM-mode is otherwise also identifiable from the Eigenmode field patterns --- or \emph{standing waves} --- and the surface current vector directions on the girders.

\subsubsection{Generating the equivalent IVU circuit model}
Based on inspection of the internal structure of the IVU, it soon became clear that at least for VHF frequencies (critical IVU resonances) the two girders inside the IVU chamber constitute a balanced transmission line. The surrounding chamber wall acts as a shield. In simpler words: two conductors inside a cylindrical enclosure with cross-sectional symmetry (see for example Ref.~\cite{Sarbacher1943} p.104 and p.117). 

The vertical struts which serve as a means for mechanical adjustment of the IVU center gap between the two girders, constitute a so-called periodic shunt loading of the balanced transmission line formed by the girders. Electrically (\gls{RF}), the struts are in fact end-shorted coaxial transmission line stubs. Their relatively short length gives an inductive character to the loading, up to a frequency where their electrical length is a quarterwave long. The inductive loading makes the balanced line in the form of two girders electrically longer than it would otherwise be. 
\\
The balanced transmission line formed by the two girders is somewhat extended and short-circuited at both ends by the flexible tapers. Due to the nature of their geometry, the taper sections also have a higher characteristic impedance than the girder sections.  

\begin{figure*}[th!]
\centering
\includegraphics[width=0.5\textwidth]{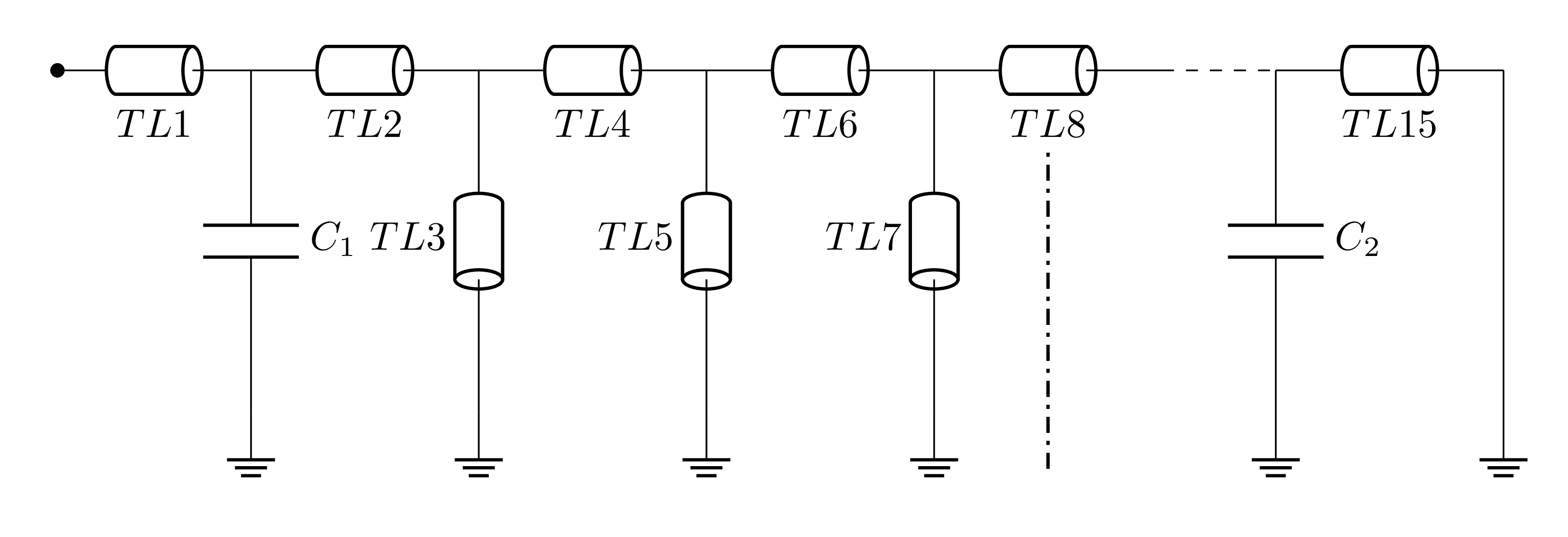}
\caption{Basic IVU EQC. Symmetry line at TL8.}
\label{fig: IVU_EQC}
\end{figure*}

The shunt capacitances $C_{1}$, $C_{2}$ in Fig.~\ref{fig: IVU_EQC} account for E-field concentrations at the abrupt and sharp ends of the girders. 
\\
With these educated inspection-based assumptions, a first \gls{EQC} was drawn up (Fig.~\ref{fig: IVU_EQC}). Using the symmetry feature of the shielded balanced transmission line mentioned above, the circuit model is an unbalanced transmission line circuit with properly adjusted transmission line characteristic impedance parameters ($Z_0^{\text{unbal}} = 0.5 Z_0^{\text{bal}}$). 

\begin{figure}
    \centering
    \includegraphics[width=0.5\linewidth]{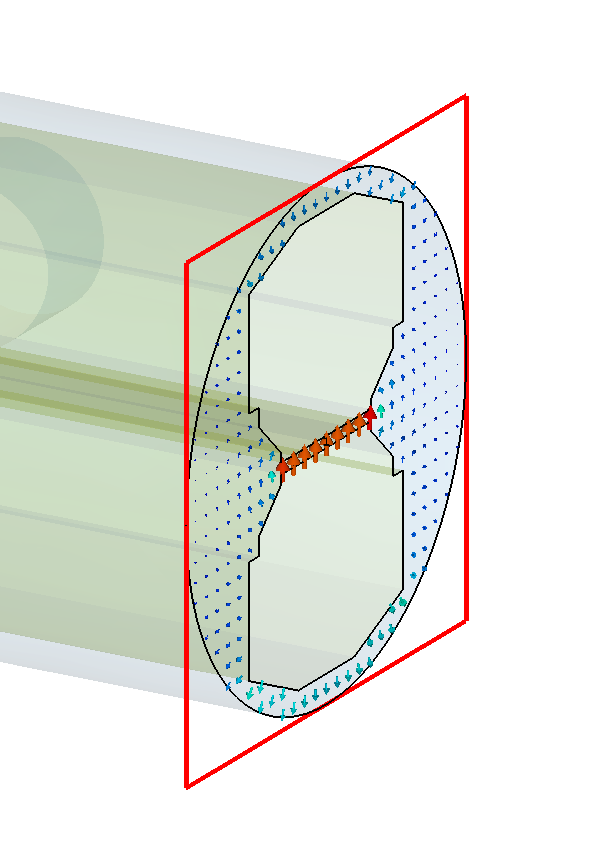}
    \caption{IVU cross-section of 2D port simulation (CST) for obtaining characteristic impedance data.}
    \label{fig:IVU cross sec}
\end{figure}

A uniform transmission line is defined by its characteristic impedance, electrical length and attenuation constant. All of these parameters must be determined for the circuit model using the results of 2D port-only EM simulation (Fig.~\ref{fig:IVU cross sec}). The \gls{3D EM} model has been suitably cut so that the correct cross-section is exposed. 

\begin{figure}
    \centering
    \includegraphics[width=1\linewidth]{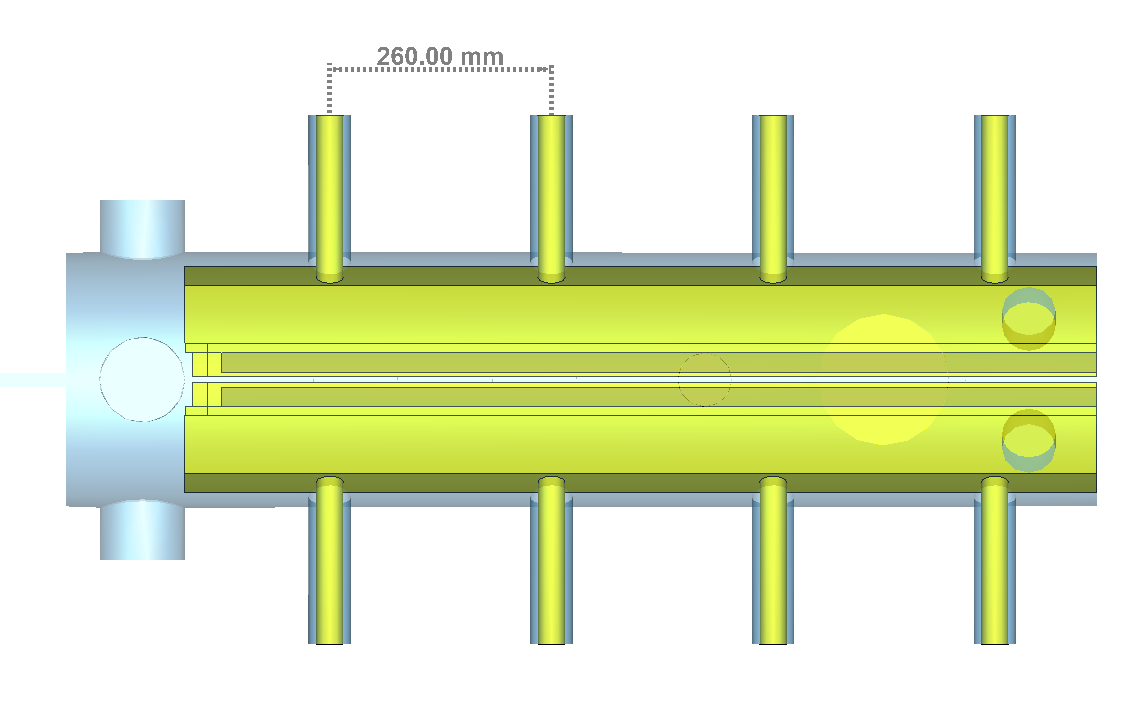}
    \caption{IVU balanced transmission line section length.}
    \label{fig:IVU balanced}
\end{figure}

The physical length dimensions of the \gls{3D EM} model are sufficient as starting values for the electrical lengths of uniform transmission line elements in the \gls{EQC} (Fig.~\ref{fig:IVU balanced}). The characteristic impedance of the coaxial line stubs formed by the struts is calculated directly from the involved radii. 
\\
In terms of the fields along the girders, the girder-strut connections represent discontinuities causing EM fields to locally divert from the pure TEM-mode. Therefore, further refinement of the relevant circuit model details is necessary. In this context a "driven IVU \gls{3D EM} model" can be useful. Such a model contains only the two periodically loaded girder-struts objects without tapers and the surrounding chamber. See Fig.~\ref{fig:IVU cross sec}. 
\\


\subsubsection{Parameters of the equivalent IVU circuit model}
Due to the rather extreme cross-sectional dimensions, the balanced-mode characteristic impedance is relatively low (around 14 ...15 Ohms, dep. on the gap size). The unbalanced equivalent circuit shown in Fig.~\ref{fig: IVU_EQC} therefore uses exactly half of that impedance. 
The initial length parameters of the circuit model require refinement because of the difference between physical length dimensions and effective electrical lengths. This difference is caused by the aforementioned local field distortions at the various junctions in the structure.

After starting values for the \gls{EQC} elements have been established as described earlier in this article, the parameter values of a valid \gls{EQC} can be found by mathematical minimisation of the differences between the Eigenmode frequencies of the \gls{3D EM} model and the circuit model represented by
\begin{equation}
\label{eqn x_i IVU}
x = \arg\min \left| f_{\text{res}_i}^{\text{EQC}}(x) - f_{\text{res}_i}^{\text{EM}} \right|
\end{equation}

Circuit model analysis does not directly yield the eigenmode frequencies. Therefore, a suitable way to find them needs to be developed.
One relatively simple way is to ensure that the circuit model uses lossy elements. A very low-impedance $S$-parameter port ($Z < 0.1$~Ohm) is then added to the circuit model input. This enables observation of an input reflection coefficient $S_{11}$ without significantly affecting the electrical properties of the overall \gls{EQC}. 
Given that each circuit resonance will consume some of the applied energy, the reflected power must be lower than the incident power. Therefore, the resonances manifest themselves as sharp peaks in the $S_{11}$ simulation result response. 

The implementation of equation (\ref{eqn: goal S_ij}) in a modern circuit simulator as an optimisation goal can for example be done by exploiting the dB({$S_{11}$}) response for each resonance. 
A circuit model simulation that yields the same resonances as those found from \gls{3D EM} simulation, while having a very close relationship to the 3D structure must by necessity be a valid \gls{EQC} for the given purpose of confirming the nature of the resonance modes (TEM). 
\\
\subsubsection{Limitations of the IVU circuit model}
It needs to be mentioned here that further analysis of our IVU \gls{EQC} validity was carried out since~\cite{Ericson2025} was published. 
It showed that the relatively strong field distortions on the girders caused by the strut connections and presence of higher modes put certain limitations on the validity of the found \gls{EQC}. 
Insofar, representation of the girder sections by uniform transmission lines --- while sufficient for an upper frequency limit of about 250~MHz --- may not yield satisfactory results for resonances above 250~MHz. Moreover, equivalent circuits with large validity bandwidth are in general very difficult to generate. 

\subsection{Linear accelerator equivalent\\circuit}
\subsubsection{General description}
\begin{figure*}[ht]
    \centering
    \includegraphics[width=0.7\linewidth]{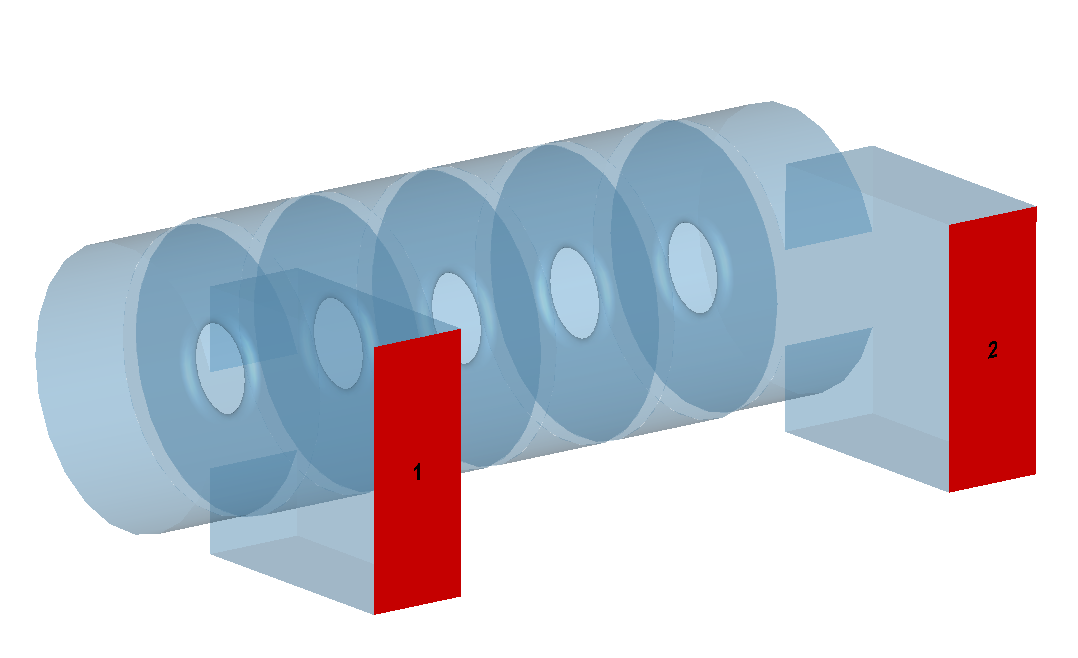}
    \caption{6-cavity linear accelerator 3D EM model.}
    \label{fig: 6-cavity linac}
\end{figure*}

As the third example, we present a highly accurate equivalent circuit for a basic 6-cavity linear accelerator, driven via a TE-10 mode \gls{WG}. As in \ref{RFPG} the narrowband nature of the linear accelerator structure makes it easy to develop an accurate \gls{EQC}.
Fundamentally, the accelerator structure shown in Fig.~\ref{fig: 6-cavity linac} has a very close relationship to a coupled-resonator bandpass filter. Therefore, known equivalent circuit techniques used in \gls{RF} engineering can be used \cite{Cameron2007}. It is of course understood that given that the structure is a chain of identical iris-coupled cavities, it is also a so-called periodic structure.

\subsubsection{Generating the linear accelerator circuit model}

The symmetry of the accelerator structure simplifies the job of generating a valid equivalent circuit. Inspection of the 3D structure shown in Fig.~\ref{fig: 6-cavity linac} shows that the functional elements for the circuit model to be determined are:

\gls{WG} - coupling-01 - res-1 - coupling-12 - res-2 - coupling-23 - res-3 - coupling-34.

The remaining elements are identical to those identified above (coupling 01 = coupling 67, Res1 = Res6, coupling 12 = coupling 56, etc.).

The coupling from the \gls{WG} to cavity-1 is established via a small intersection of the \gls{WG} and the cavity. It is inductive in nature because the opening to the cavity at the end of the \gls{WG} has the form of an inductive iris~\cite{Marcuvitz1951} and therefore, a simple mutual coupling between an inductor representing the \gls{WG} iris and the inductance of a parallel-tuned circuit for resonator-1 is appropriate and sufficient. The remainder of the \gls{EQC} is marked by capacitively (E-field) coupled cavities represented by parallel-tuned LC resonant circuits. 

\begin{figure*}[ht]
\centering
\includegraphics[width=0.5\textwidth]{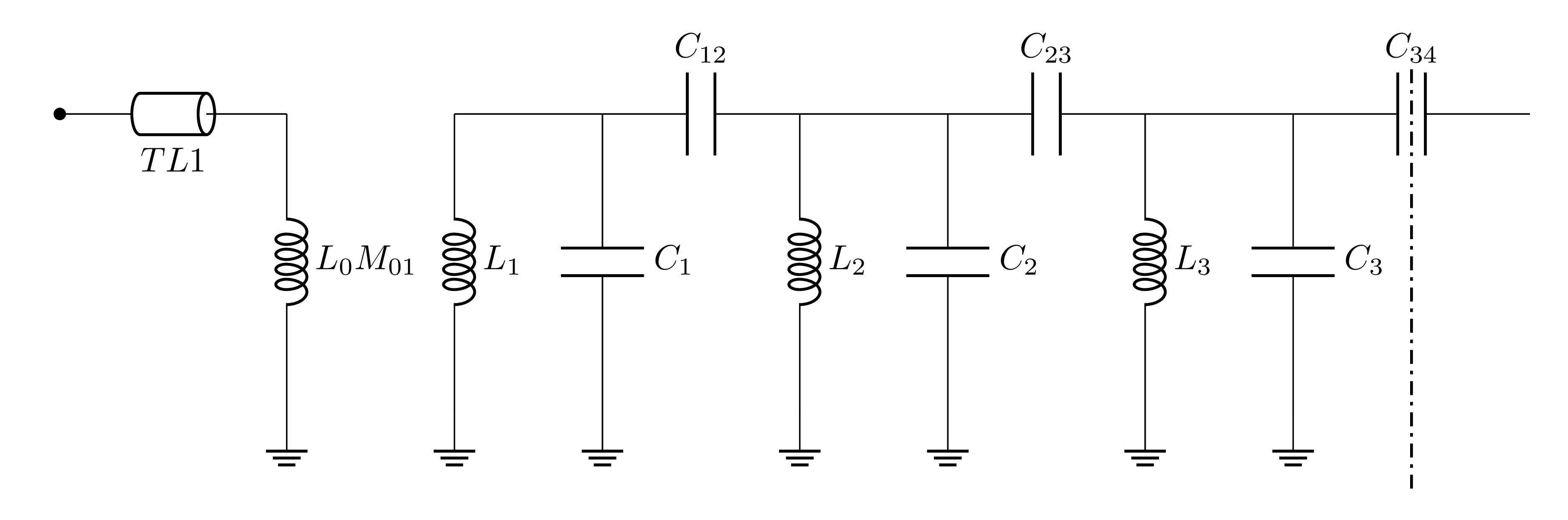}
\caption{EQC of a 6-cavity linear accelerator. Only left symmetrical half shown.}
\label{fig: linac_eqc}
\end{figure*}

Fig.\ref{fig: linac_eqc} shows the \gls{EQC} of the linear accelerator as depicted in Fig.~\ref{fig: 6-cavity linac}. The sufficiently equivalent circuit element for the rectangular input and output (\gls{WG}) line sections is a simple uniform transmission line element with the characteristic impedance of the \gls{WG} at band centre. The initial electrical length of the \gls{WG} has been obtained from standard \gls{WG} equations. The narrowband nature of the structure permits us to neglect the non-linear frequency-dependency of the \gls{WG} wavelength. 

Similar to the earlier \gls{RF} particle gun example (\ref{RFPG}), the cavities of the linear accelerator are represented as simple parallel-tuned $LC$ resonators with capacitive (E-field) coupling between them.

\begin{figure}[ht]
    \centering
    \includegraphics[width=1\linewidth]{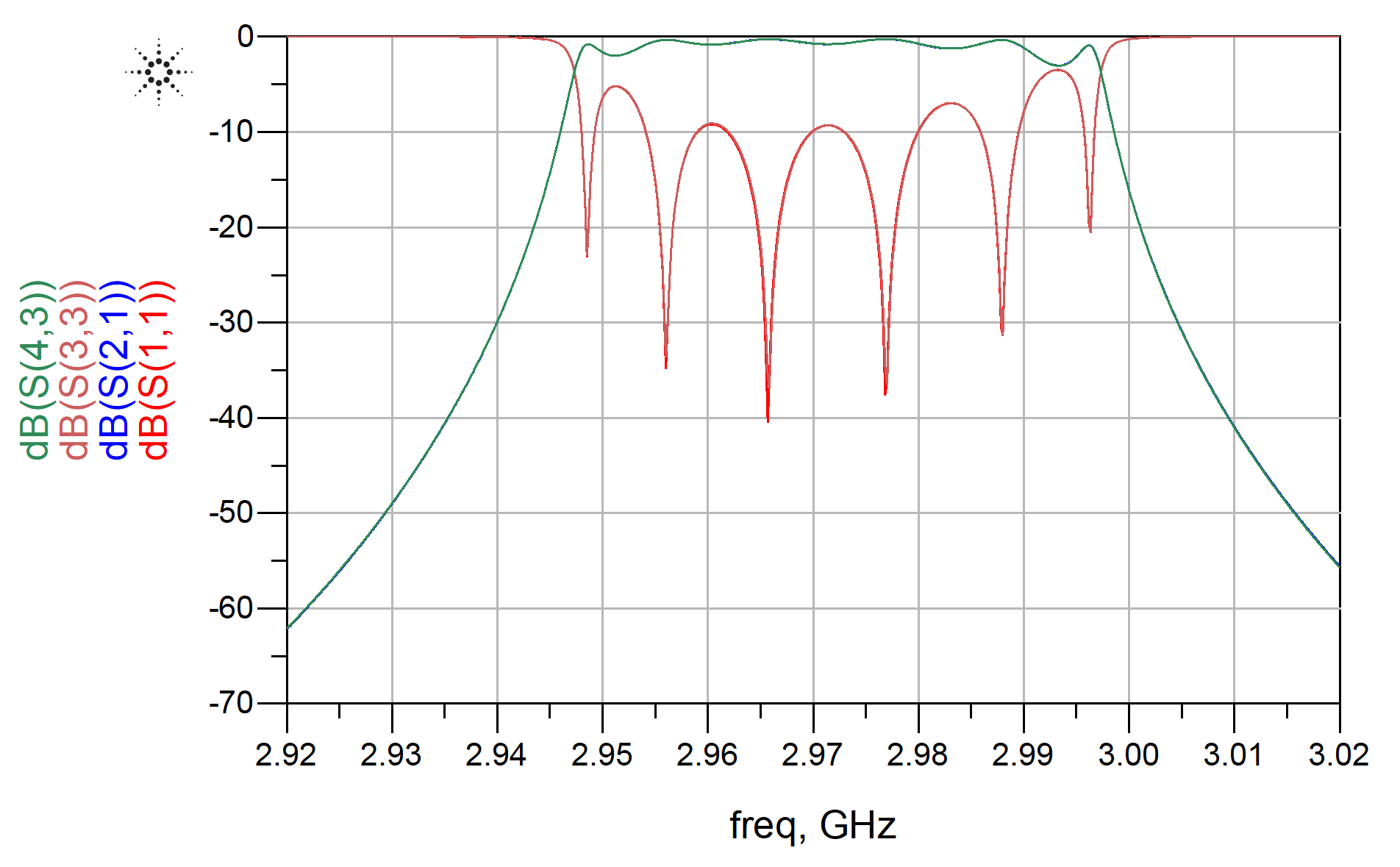}
    \caption{Equivalent circuit and 3D EM $S$-parameter simulation results.}
    \label{fig: 6-cav linac dBSpar}
\end{figure}

\begin{figure}[ht]
    \centering
    \includegraphics[width=1\linewidth]{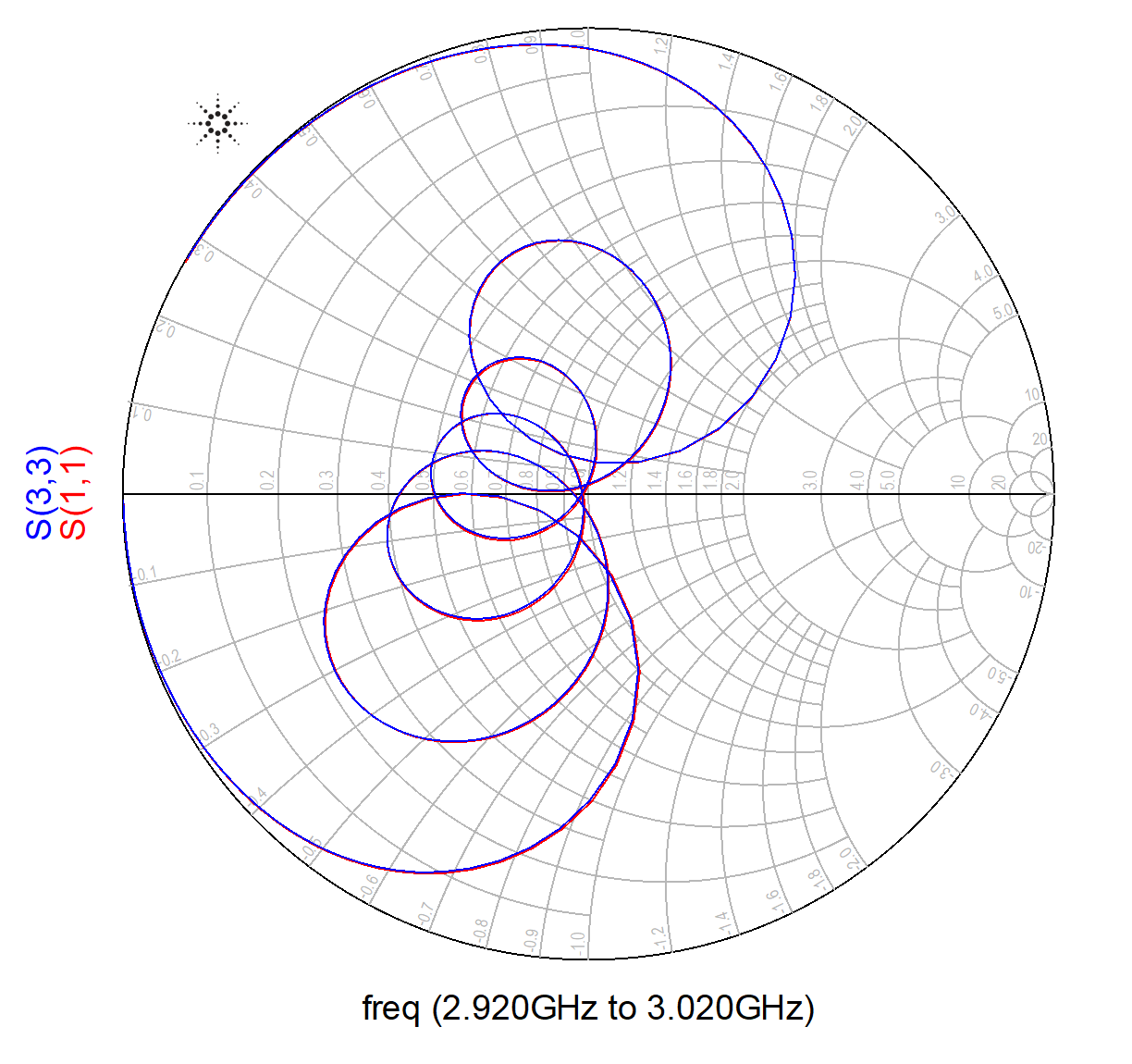}
    \caption{\gls{EQC} and \gls{3D EM} {$S_{11}$} simulation results in polar coordinates.}
    \label{fig: 6-cav linac s11}
\end{figure}

Fig.~\ref{fig: 6-cav linac dBSpar} represents an $S$-parameter deciBel magnitude comparison. The \gls{EQC} tracks the \gls{3D EM} response very well down to less than $-60$~dB in the $S_{21}$ response.
In order to illustrate that the \gls{EQC} produces both the $S$-parameter magnitude and phase-angle with highest accuracy with reference to the \gls{3D EM} model response, we include the complex $S_{11}$ polar coordinates graph in Fig.~\ref{fig: 6-cav linac s11}. 

With reference to Fig.~\ref{fig: linac_eqc} the \gls{EQC} element values are listed in Table~\ref{tab:tans_param} 
\begin{table}[h]
\caption{Values for the linac model EQC in Fig.~\ref{fig: linac_eqc}.}
\label{tab:tans_param}
\centering
\begin{tabular}{|l|c|c|c|}
\hline
\textbf{Component} & \textbf{Parameter} & \textbf{Value} & \textbf{Unit} \\
\hline
\multirow{2}{*}{TL1} & $Z_0$ & 520 & Ohm \\
                     & $l$ & 103.18 & mm \\
\hline
\multirow{6}{*}{Res1} & $M_\text{01}$ & 0.212 & nH \\
                      & $L_\text{0}$ & 0.0 & nH \\
                      & $f_{\text{res1}}$ & 2971.12 & MHz \\     
                      & $L_1$ & 0.2118 & nH \\
                      & $C_1$ & 13.471 & pF \\
                      & $C_{12}$ & 0.0768 & pF \\
\hline
\multirow{4}{*}{Res2} & $f_{\text{res2}}$ & 2972.03 & MHz \\
                      & $L_2$ & 0.6523 & nH \\
                      & $C_2$ & 4.278 & pF \\
                      & $C_{23}$ & 0.0413 & pF \\
\hline
\multirow{4}{*}{Res3} & $f_{\text{res3}}$ & 2972.12 & MHz \\
                      & $L_3$ & 0.6074 & nH \\
                      & $C_3$ & 4.637 & pF \\
                      & $C_{34}$ & 0.0428 & pF \\
\hline
\multicolumn{2}{|l|}{$Q_{\text{res1,6}}$} & 12408 & -- \\
\multicolumn{2}{|l|}{$Q_{\text{res2,3,4,5}}$} & 11861 & -- \\
\hline
\end{tabular}
\end{table}
The remaining element values (Res4 to Res6) are given by the symmetry of the structure ($C_1 = C_6$ etc.). 
The inductance values of the resonant circuits were obtained via the given resonant frequencies and the nodal capacitances as described earlier Eq.~(\ref{eqn: L calc}). 
Note, that for the end-resonators, the above given unloaded resonator $Q$-factors are not the \emph{operating Q-factors}. The latter result from the resistive loading due to the port-couplings. (The zero value of L0 is due to the implementation of coupled inductances in the used circuit simulator).

The \gls{EQC} element values directly tell us that the end-resonators have a significantly lower equivalent resonator inductance ($L_1$, $L_6$). The reason for this is twofold: a) absence of a second coupling iris b) connection to \gls{WG}. Similarly, the end resonators exhibit a slightly higher unloaded-Q due the absence of a second iris hole (lower surface current density). Due to the relatively strong coupling to a port, both end-resonators operate under a so-called \emph{loaded Q-factor}, meaning that they are insofar very different to the inner resonators. The loaded-$Q$ is usually much lower than the unloaded-$Q$. It can be determined via \gls{3D EM} Eigenmode simulation of an end-resonator with attached \gls{WG}. 

Given the strong similarity between the accelerator structure and a coupled-resonator bandpass filter, the so-called \emph{coupling matrix} approach could also be used to analyse the equivalent circuit \cite{Cameron2007}.

\subsection{Crab cavity equivalent circuit model}
\subsubsection{General description}
With reference to \cite{Apsimon2017} (CEBAF, Table 1) a $LC$ equivalent circuit for a 4-rod deflecting crab cavity was derived using the method presented in this article. The symmetry of the structure allows us to restrict the \gls{3D EM} model to just two coupled end-loaded resonator rods inside a cavity of given dimensions. Realising that the "LOM" and \emph{Deflecting} modes are in fact the normal and well-known even- and odd-mode resonances of a pair of coupled (TEM-mode) resonators, the equivalent $LC$ circuit has the standard topology used for two coupled resonators (see Ref.~\cite{Cameron2007} Ch.~14.2). 
\subsubsection{Generating the crab cavity \gls{EQC}}
Given the relatively short length of the rods relative to a quarter wavelength, magnetic (inductive) coupling between the two resonators must exist. 
Much of the electric field of the resonator rods is concentrated in the end-loading capacitance that exists between the open-ends of the rods and the adjacent wall. 
However, a small capacitance also exists between the open ends of the resonator rods. This capacitance is somewhat counter-productive to the dominant inductive coupling and therefore it reduces the magnetic field coupling between the resonator rods. For larger separation between the rods ($2S$ in \cite{Apsimon2017}) this capacitance quickly reduces while the end-loading capacitances of the rods increase (Fig.~\ref{fig: Crab_EQC_C1_3}). 

\begin{figure*}[ht]
\centering
\includegraphics[width=0.5\textwidth]{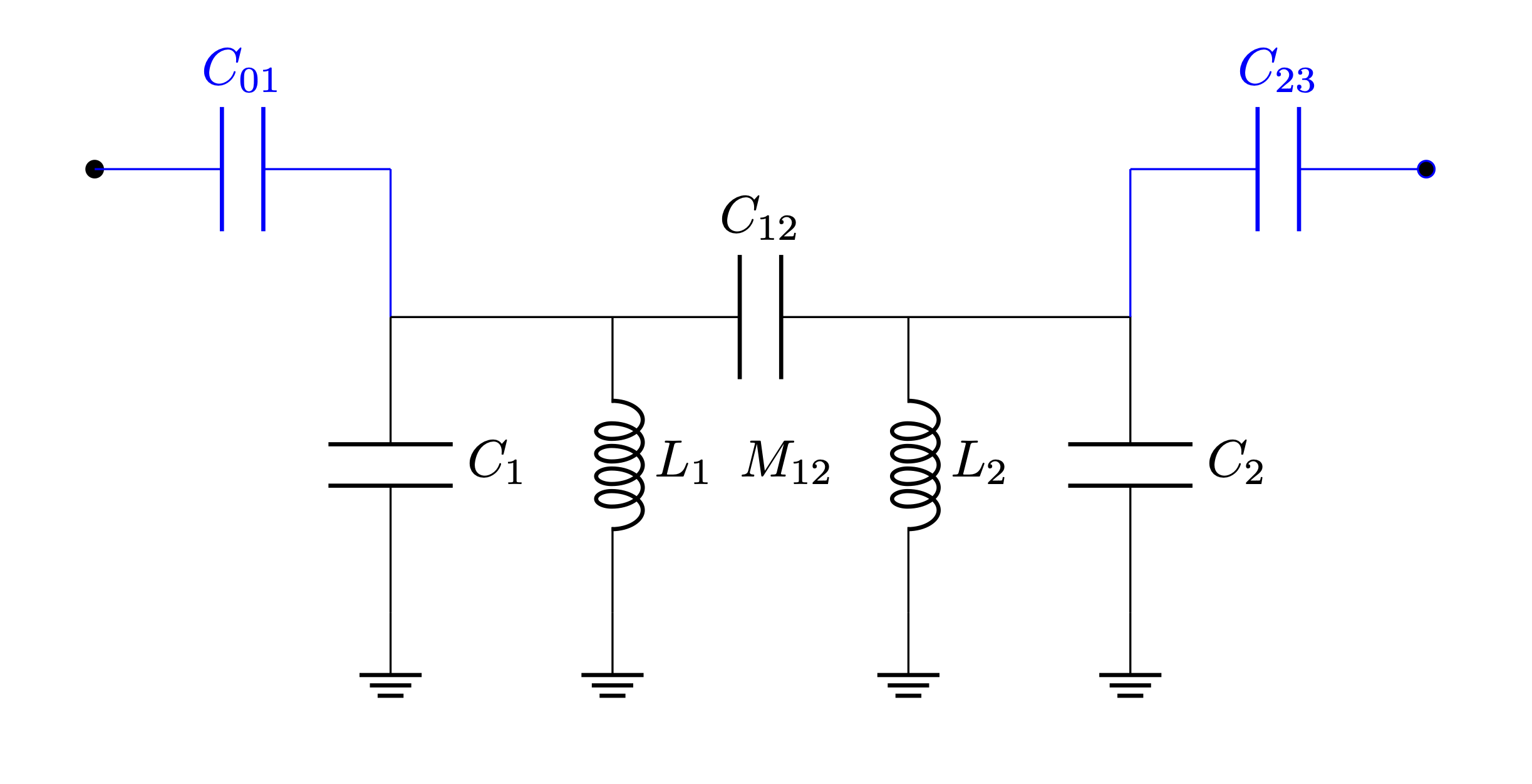}
\caption{Crab cavity EQC. Blue: Very weak probe capacitances $C_{01}$, $C_{23} \ll C_1$, $C_2$, $C_{12}$ (see text).}
\label{fig: Crab_EQC}
\end{figure*}

\begin{figure}[ht]
    \centering
    \includegraphics[width=1\linewidth]{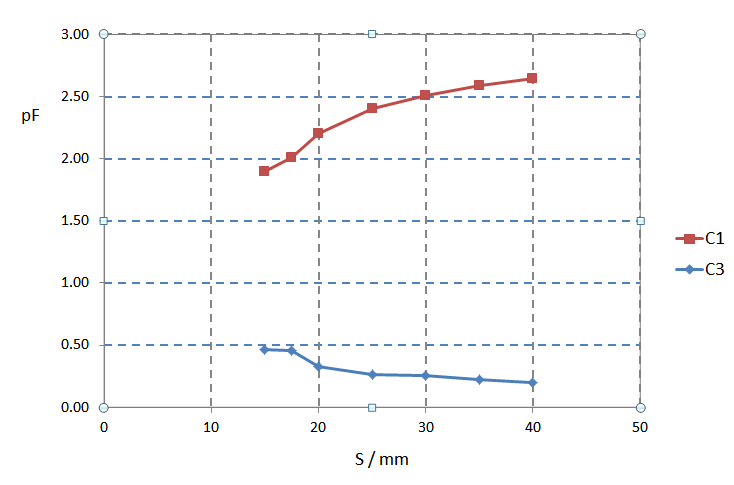}
    \caption{Equivalent $LC$ Resonator capacitances $C1,2$ and $C_3$ versus rod half-spacing S (Fig.~\ref{fig: Crab_EQC}} in Ref.~\cite{Apsimon2017}.)
    \label{fig: Crab_EQC_C1_3}
\end{figure}

With reference to paragraph-\ref{Generating EQC} of this article and Ref.~\cite{Collins2001,Sarbacher1943} we can use analytical equations to find starting values for the equivalent $LC$ circuit element values for the resonators. 
The remaining starting values for the coupling elements are found by systematic trials under observation of the transmission loss graph ($S_{21}$). 
Refinement of circuit element values follows as described in Sec.~\ref{sec:refine} of this article. 
The \gls{EQC} is shown in Fig.~\ref{fig: Crab_EQC}. 

\begin{figure}[ht]
    \centering
    \includegraphics[width=1\linewidth]{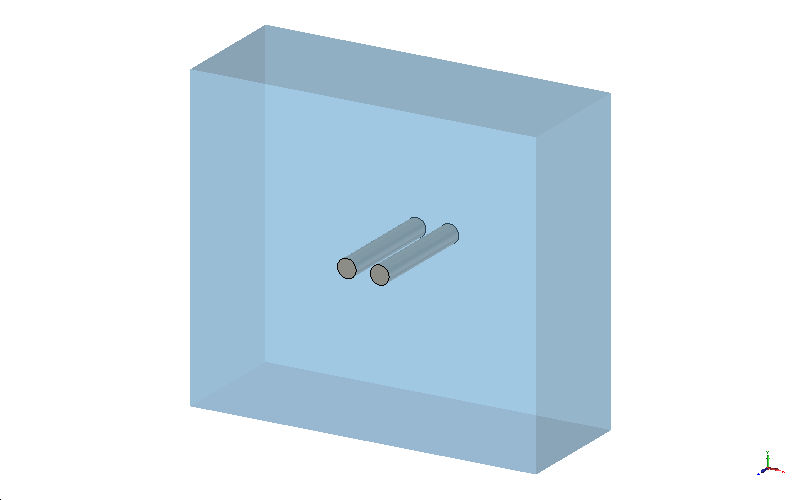}
    \caption{3D model view of half Crab cavity.}
    \label{fig: Crab_3D}
\end{figure}

\begin{figure}[ht]
    \centering
    \includegraphics[width=1\linewidth]{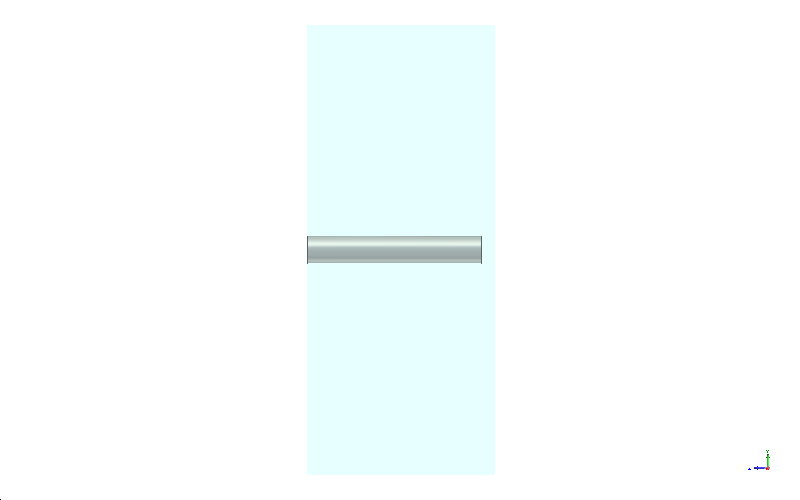}
    \caption{2D view of half Crab cavity.}
    \label{fig: Crab_2D}
\end{figure}

\begin{figure}[ht]
    \centering
    \includegraphics[width=1\linewidth]{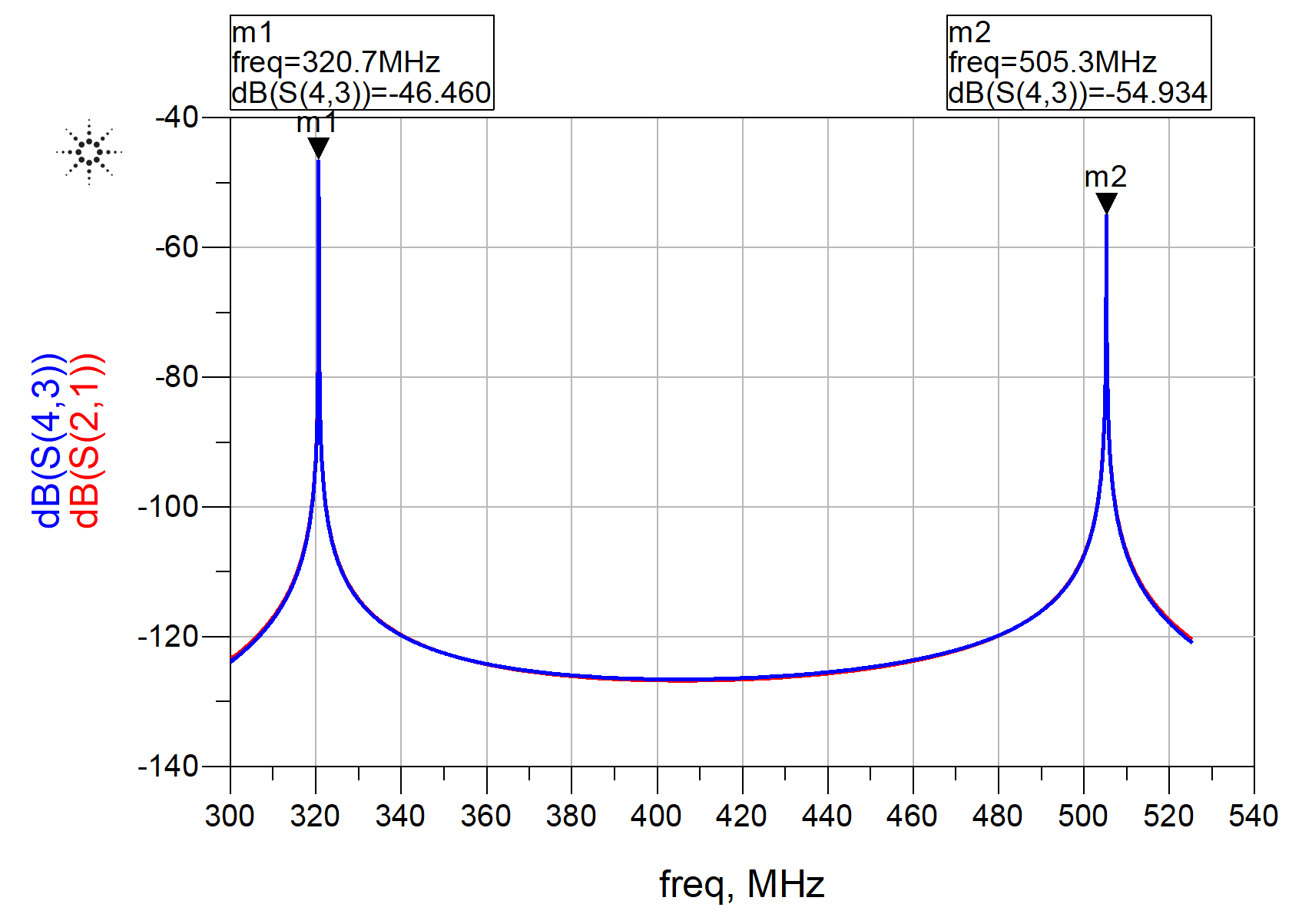}
    \caption{Simulation results for S = 17.5~mm showing near perfect agreement between 3D EM (CST) model and \gls{EQC}.}
    \label{fig: Crab_EQC_graph}
\end{figure}

For reasons of simplicity, both the \gls{3D EM} and the circuit model use the so-called \emph{driven mode} which directly yields both even- and odd-mode resonance peaks in the $S_{21}$ response graph (Fig.~\ref{fig: Crab_EQC_graph}). 
For this purpose very small capacitive couplings were added to the structure to obtain just a small peak transmission of about $-50$~dB (Fig.~\ref{fig: Crab_EQC_graph}). 
The vanishingly small additional couplings to the resonator rods therefore fully preserve the electromagnetic properties of the overall structure. 
In the circuit model, these input and output couplings are represented by the capacitances $C_{01}$ and $C_{23}$ in Fig.~\ref{fig: Crab_EQC}.

Due to the relatively strong (magnetic field) coupling between the resonator rods, the even- and odd-mode resonances are quite far apart. It is important to note that the two lowest order natural modes of a pair of coupled resonators are entirely a direct consequence of the coupling between them~\cite{Cameron2007} (Ch. 14.2).

With reference to Fig.~\ref{fig: Crab_EQC}, the values of the circuit elements for $S = 17.5$~mm~\cite{Apsimon2017} ($2S$ = rod spacing) are listed in Table~\ref{tab:crab_param}.
\begin{table}[ht]
\caption{Values of circuit components for Fig.~\ref{fig: Crab_EQC_C1_3}.}
\label{tab:crab_param}
\centering
\begin{tabular}{|l|c|c|}
\hline
\textbf{Parameter} & \textbf{Value} & \textbf{Unit} \\
\hline
Port Impedance & 117 & Ohm \\
$C_{01} = C_{23}$ & 0.0009 & pF \\
$L_1 = L_2$ & 74.53 & nH \\
$C_1 = C_2$ & 2.01 & pF \\
$C_{12}$ & 0.457 & pF \\
$M_{12}$ & 42.67 & nH \\
\hline
\end{tabular}
\end{table}

The natural frequency dependency of the mutual inductance $M_{12}$ was considered with a linear frequency slope factor of 0.372 (slope zero crossing at 478.96~MHz). 

By repeatedly using the same process described above, the relationship between the parameters of the 3D structure and the equivalent circuit element values can be established. 
Fig.~\ref{fig: Crab_EQC_C1_3} serves as an example showing the relationship between the rod half-spacing, $S$ and the equivalent circuit capacitances $C_1$, $C_2$ and $C_3$. 
With reference to \cite{Apsimon2017} it is noted that the equivalent circuit in Fig.~\ref{fig: Crab_EQC} is valid as is for any given \gls{3D EM} $S$-parameter data set and does not require any form of circuit topology or other modifications. 
Furthermore, the circuit elements have a direct relation to the 3D model and its various dimensions. 

\section{Applications of equivalent circuits}
\label{App}
A valid equivalent circuit can serve a wide range of purposes. The list below provides information on typical applications.

\begin{enumerate}
    \item \textbf{Tuning status information}

    For tuned structures, like accelerators or particle guns, equivalent circuit simulation instantly reveals the tuning status of the \gls{3D EM} model. Mistuned cavities are directly identifiable.
    \item \textbf{Quick answers to "what-if" questions} 
    
    Effect of changes in certain circuit element parameters which have a strong relation to certain dimensions of the \gls{3D EM} model.
    \item \textbf{Diagnostics of pathological structures}
    
    A problem in the \gls{3D EM} model may become visible immediately in the values of the equivalent circuit elements.
    \item \textbf{Instant observation of the effect of element value changes}
    
     The effect of \gls{EQC} element value changes can be seen quasi immediately in the resulting change in the response graphs. On the basis of the gained information a certain well-targeted change can then be carefully implemented at the appropriate related location in the \gls{3D EM} model with prior knowledge of the resulting effect.
     
     \item \textbf{Fast optimisation of \gls{EQC}}
     
     A certain desired design goal can be applied to the EQC in the context of an optimisation of the circuit model element values. The post-optimisation circuit model element values (in relation to the pre-optimisation values) can directly point to required changes in the \gls{3D EM} model. In certain cases the performance potential of a given structure may be revealed through circuit simulation. As always, it is mandatory to apply appropriate constraints to all circuit element values during optimisations in order to ensure that the required changes in the \gls{3D EM} model are realisable.
     
     \item \textbf{Nodal voltage analysis} 
     
     Besides $S$-parameter analysis of the \gls{EQC}, nodal voltage and branch current analysis is also possible. In the case of accelerating structures for example, resonator-voltage relates to the E-fieldstrength in a cavity. Therefore, nodal voltage analysis offers additional design and diagnostic value. 
     It must be mentioned that, in certain cases, due to the presence of internal impedance transforming features of the \gls{EQC}, the absolute nodal voltages may not always be highly accurate. 
\end{enumerate}

It is advisable to first practice deriving and generating \gls{EQC}s of relatively simple structures with known and well-understood properties before proceeding to more complex structures. 

\section{Conclusion}
The process of deriving and generating equivalent circuits for complex structures was presented in this paper. Four representative examples illustrate and explain the derivation of an equivalent circuit from \gls{3D EM} simulation result data. Important details of the process were highlighted and explained to enable readers to use the process without encountering major obstacles. The immense usefulness of \gls{EQC}s is illustrated by a list of numerous applications. 

Rather than a \emph{3D EM only} approach, it is very beneficial to synchronise the \gls{3D EM} simulation model with an accurate \gls{EQC} and use both models in tandem \cite{Levy2000,Apsimon2017}. 

\section{Note}
For reasons of clarity and readability, the equivalent circuits in this paper show only the circuit elements and their names. The element values are given within this article in table form in the related text. For the same reasons, the loss resistors are not shown in the schematics. Their values follow directly from the given quality factor and the resonator $LC$ values.
Further information can be obtained from the authors upon request.

\printglossary

\bibliography{ref}
\bibliographystyle{unsrt}

\end{document}